\documentclass[twocolumn]{aastex62}

\usepackage{times}
\usepackage{graphicx}

\graphicspath{{./}{figures/}}

\accepted{to ApJL}

\shorttitle{J1010+1413: empirical constraints on the GWB}
\shortauthors{Goulding et al.}

\newcommand{\Msun}{$M_{\odot}$}
\newcommand{\Lsun}{$L_{\odot}$}

\begin{document}

\title{Discovery of a close-separation binary quasar at the heart of a $z \sim 0.2$ merging galaxy and its implications for low-frequency gravitational waves}

\correspondingauthor{Andy D. Goulding}
\email{goulding@astro.princeton.edu}

\author{Andy D. Goulding}
\affil{Department of Astrophysical Sciences, Princeton University, Princeton, NJ 08540, USA}

\author{Kris Pardo}
\affil{Department of Astrophysical Sciences, Princeton University, Princeton, NJ 08540, USA}

\author{Jenny E. Greene}
\affil{Department of Astrophysical Sciences, Princeton University, Princeton, NJ 08540, USA}

\author{Chiara M. F. Mingarelli}
\affil{Flatiron Institute, 162 Fifth Ave, New York, NY 10010, USA.}

\author{Kristina Nyland}
\affil{National Research Council, resident at the U.S. Naval Research Laboratory 4555 Overlook Ave. SW,
Washington, DC 20375, USA.}

\author{Michael A. Strauss}
\affil{Department of Astrophysical Sciences, Princeton University, Princeton, NJ 08540, USA}

\begin{abstract}
Supermassive black hole (SMBH) binaries with masses of $\sim 10^8$--$10^9$\Msun\ are expected to  dominate the contribution to the as-yet undetected gravitational wave background (GWB) signal at the nanohertz frequencies accessible to Pulsar Timing Arrays (PTA). We currently lack firm empirical constraints on the amplitude of the GWB due to the dearth of confirmed SMBH binaries in the required mass range. Using HST/WFC3 images, we have discovered a $z \sim 0.2$ quasar hosted in a merger remnant with two closely separated (0.13$''$ or $\sim$430pc) continuum cores at the heart of the galaxy SDSSJ1010+1413. The two cores are spatially coincident with two powerful [OIII]-emitting point sources with quasar-like luminosities ($L_{\rm AGN} \sim 5 \times 10^{46}$~erg~s$^{-1}$), suggesting the presence of a bound SMBH system, each with $M_{\rm BH} > 4 \times 10^8$\Msun. We place an upper limit on the merging timescale of the SMBH pair of 2.5 billion years, roughly the Universe lookback time at $z \sim 0.2$. There is likely a population of quasar binaries similar to SDSSJ1010+1413 that contribute to a stochastic GWB that should be detected in the next several years. If the GWB is not detected this could indicate that SMBHs merge only over extremely long timescales, remaining as close separation binaries for many Hubble times, the so-called `final-parsec problem'. 
\end{abstract}


\section{Introduction}

Cosmological models of structure formation predict that galaxies undergo frequent mergers throughout their history \citep{Volonteri03,Springel05}. SMBHs from each progenitor quickly sink towards the central 0.1--1~kpc region of the merger remnant due to dynamical friction. This SMBH pair may eventually form a bound binary system capable of emitting gravitational waves (GWs) before final coalescence \citep{Begelman1980}. SMBH binaries with masses of $M_{\rm BH} \approx 10^8$--$10^9$\Msun\ are expected to comprise the dominant contribution to the as-yet undetected GW background (GWB) signal at the nanohertz frequencies accessible to pulsar timing arrays (PTAs; \citealt{Sesana2008,btc+18}). The current theoretical predictions on the precise amplitude and composition of the GWB vary dramatically, and are limited, in part, by the lack of overall empirical constraints on the occurrence of high-mass SMBH pairs.

\begin{figure*}
    \centering
    \includegraphics[width=\textwidth]{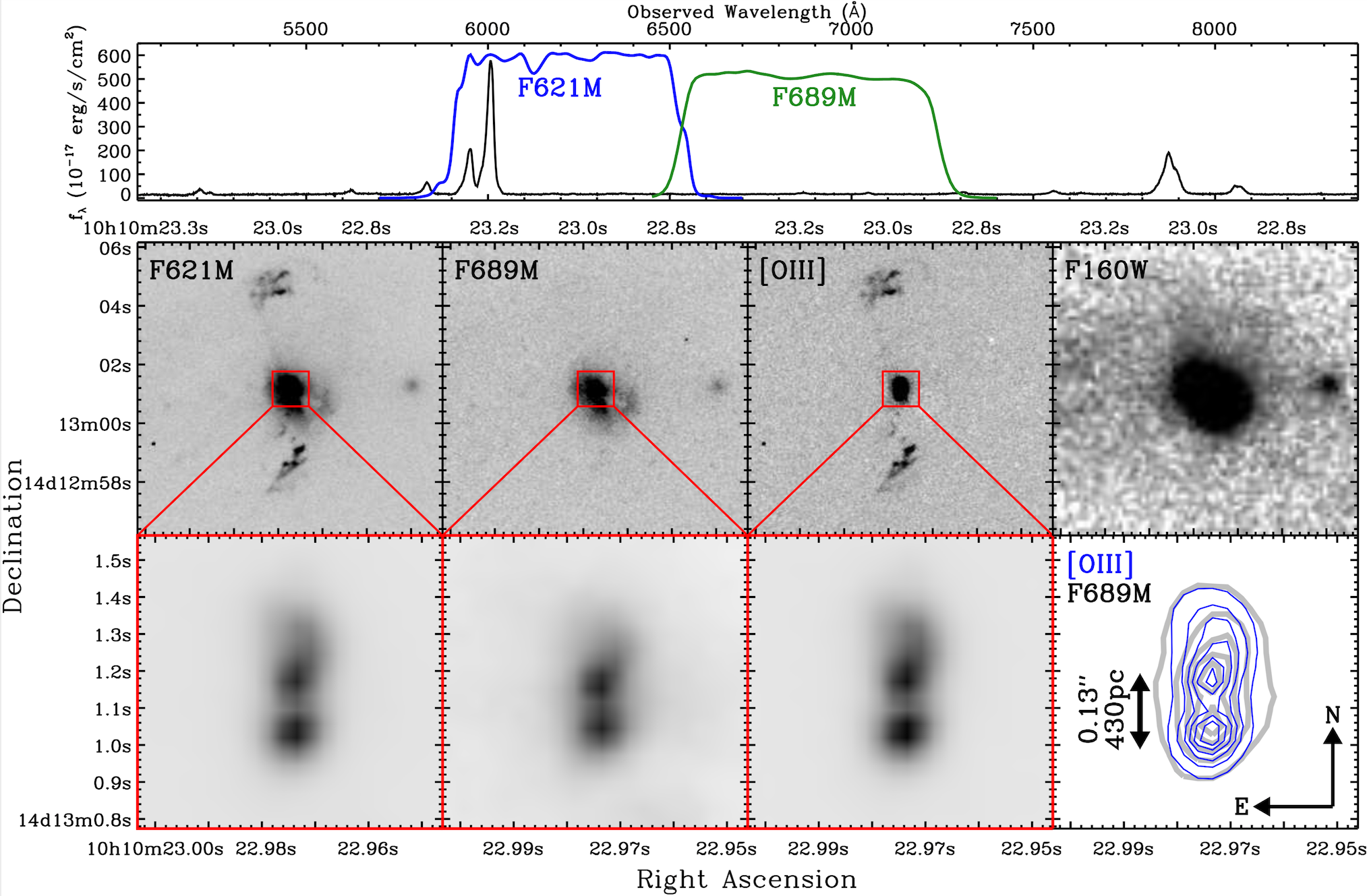}
    \caption{Analysis of the central region of J1010+1413. Upper panel: SDSS 2$''$ fiber spectrum, overlaid are the WFC3/UVIS transmission curves for the F612M (blue) and F689M (green) filters. Middle panels left to right: WFC3/UVIS medium-bands F621M; F689M; [OIII]=F621M--F689M; F160W. Lower panels left to right: zoom and contrast rescaling of the middle panels. Contours of F689M continuum (black) and [OIII] images (blue) are shown in the lower-right panel. Both the F689M stellar continuum-only image and the [OIII] images clearly show two distinct point sources, suggesting two nuclei each with their own accreting SMBHs.}
    \label{fig:hst}
\end{figure*}

Emission produced through accretion acts as a useful signpost for evidence of SMBHs. Indeed, extremely high-resolution radio observations have serendipitously identified very close separation sub-parsec scale SMBH binary candidates \citep{Rodriguez2006}. While there have been several claims of binary SMBHs in some nearby galaxies (e.g., \citealt{Fabbiano2011}), these have subsequently been challenged with expanded datasets and/or improved analysis techniques (e.g., \citealt{Finlez18} and references therein). AGN surveys have identified wider-separation (several to tens kpc) SMBH-pair candidates by harnessing highly-penetrating X-rays and follow-up with ground-based near-IR adaptive optics systems \citep{Liu13,Koss18}. Due to limitations in angular resolution these investigations are typically performed at lower redshifts, where the AGN studied tend to be relatively low-luminosity ($L_{\rm AGN} < 10^{44}$~erg~s$^{-1}$). Consequently, these dual SMBH systems are predisposed to lower masses ($M_{\rm BH} \approx 10^7$\Msun) often at kiloparsec separations \citep{Liu10,Woo14}, such as the well-studied nearby merging galaxy NGC~6240, which hosts two actively growing SMBHs separated by $\sim$1~kpc \citep{Muller18}. Systems such as NGC~6240 do not appreciably contribute to the GWB that will be detected in the PTA band due to their low $M_{\rm BH}$ \citep{Sesana2008}. Much larger volumes must be searched to find the most luminous quasars, and hence most massive binary BHs, that contribute to the PTA signal. Utilizing high spatial-resolution observations with the Hubble Space Telescope (\emph{HST}), studies have begun to identify a more massive population of kpc-scale separation SMBH pairs through the detection of distinct multiple nuclear cores coincident with unresolved AGN emission \citep{Xu09,Fu12}. However, to begin to place empirical constraints on the GWB we must characterize the number of $10^8-10^9$\Msun\ SMBH pairs with sufficiently small separations that they may merge before the present day.

Here we present an observational anchor to predictions for a GWB signal that can be detected in the PTA band. SDSSJ101022.96+141300.9 (hereafter, J1010+1413) is a late-stage merging galaxy at $z \sim 0.198$ (angular scale of 3.27kpc/$''$) with high equivalent-width emission lines \citep{Mullaney13} in the Sloan Digital Sky Survey \citep[SDSS;][]{York2000}. At mid-infrared wavelengths (22$\mu m$), it is one of the most luminous systems ($L_{\rm AGN} \gtrsim 6 \times 10^{46}$~erg~s$^{-1}$) at $z \sim 0.2$ identified with the Wide-field Infrared Survey Explorer (WISE). Using high spatial resolution (0.03$''$/pixel) multi-band imaging from HST's Wide Field Camera 3 (WFC3) instrument, we identified two distinct quasar-produced [OIII]5007-emitting regions that are spatially co-incident with two nuclear stellar cores buried close to the center. These two point sources are separated by 0.13$''$, a projected separation of only 430 parsecs at the distance of J1010+1413, providing strong evidence for a central SMBH pair at the heart of this merging galaxy.

\section{Target Selection and Observations}

J1010+1413 was initially characterized as having strong asymmetries in its [OIII]5007 emission line profile as part of an in-depth study of nearby accreting SMBHs identified in SDSS \citep{Mullaney13}. It is one of the most luminous quasars at $z \sim 0.1$--0.2 based on both [OIII] ($L_{\rm [OIII],dered} \sim 1.2 \times 10^{44}$erg~s$^{-1}$) and 22$\mu$m emission. The central few kpc of J1010+1413 is strongly AGN dominated ([OIII]/H$\beta \sim 12.4$) and was previously found to be irregular and kinematically complex, with broad, spatially unresolved [OIII] and H$\beta$ emission ($W_{80} \sim 1350-1450$~km~s$^{-1}$) based upon Gemini-GMOS IFU data \citep{Harrison2014}. 

\subsection{HST medium \& broad-band imaging}
J1010+1413 was imaged on October 17 2017 with WFC3/IR in the F160W filter, and with the WFC3/UVIS instrument in two optical medium bands, F621M and F689M for a total of one orbit (Proposal:14730; PI:Goulding). Our F160W observation with a spatial resolution of 0.13$''$/pixel provides a relatively clean measure of the stellar light (rest-frame 1.3$\mu$m). An azimuthally-symmetric surface brightness (SB) profile was constructed from the F160W image to measure the stellar light of J1010+1413. The SB profile is detected to $r \sim 7.1''$ ($\sim 23$~kpc), and is dominated by inhomogeneous low SB emission beyond $r \sim 2.25''$, consistent with tidal debris from a merger event. We measure a total flux of $f_{160W} \sim 3.23\pm0.03 \times 10^{-12}$~erg~s$^{-1}$cm$^{-2}$, i.e., a total luminosity of $L_{160} \sim 1.1 \times 10^{11}$\Lsun, from which we infer a total stellar mass of $M_{*}\sim0.7$--$1.5 \times 10^{11}$\Msun. 

The WFC3/UVIS F621M observation (0.04$''$/pixel) provides a detailed image of the continuum galaxy light combined with emission from the AGN-dominated [OIII]4959+5007\AA\ emission line doublet, which is redshifted into the filter. By contrast, the F689M filter covers no significant emission lines and thus only detects emission arising from the galaxy continuum. On small scales, the F689M filter image reveals two distinct and resolved stellar-continuum point sources that are separated by 0.13$''$ in the nuclear region of J1010+1413 (Fig.~1). The amplitudes of the two nuclear cores are very similar, 1:1.2, suggesting similar stellar masses but the two cores are not uniquely resolved in the F160W image due to the lower resolution and S/N.

We subtracted the F621M from the F689M image to investigate the spatial distribution of the combined [OIII] doublet emission, and test for the presence of two distinct narrow-line regions (NLR) in the center of J1010+1413. We solved for the marginal ($<$0.02$''$) astrometric offset between the two F621M and F689M filters, and normalized the flux-calibrated images to give a net-zero sum of the extended galaxy light. Within the central kpc region (r$\lesssim$0.3$''$), we identify two distinct [OIII]-emitting regions that are spatially co-incident with the two nuclear point sources observed in the F689M continuum image. We interpret this as evidence for two remnant cores from the galaxy merger that each possess a central accreting SMBH, producing two distinct [OIII]-luminous NLRs.

Using the [OIII] image, we confirm the presence of the extended emission line region at large radial scales of $r \sim4$'' (14~kpc), the kinematics of which was previously studied in \citet{Sun2017}. Their previous Magellan IMACS spectroscopy shows that the extended [OIII] emission is kinematically cold, and not necessarily an outflow from the central galaxy. Combined with our sensitive WFC3/UVIS observations, the large-scale [OIII] appears to be emitted from gas that was tidally stripped during the galaxy merger \citep{Harrison2014}, and is now being illuminated by the central quasars. Indeed, we find evidence for two collimated emission features oriented at $\sim$11 and 342 degrees from North, consistent with the presence of two distinct accretion disks (Fig.~2).

\begin{figure*}
    \centering
    \includegraphics[width=\textwidth]{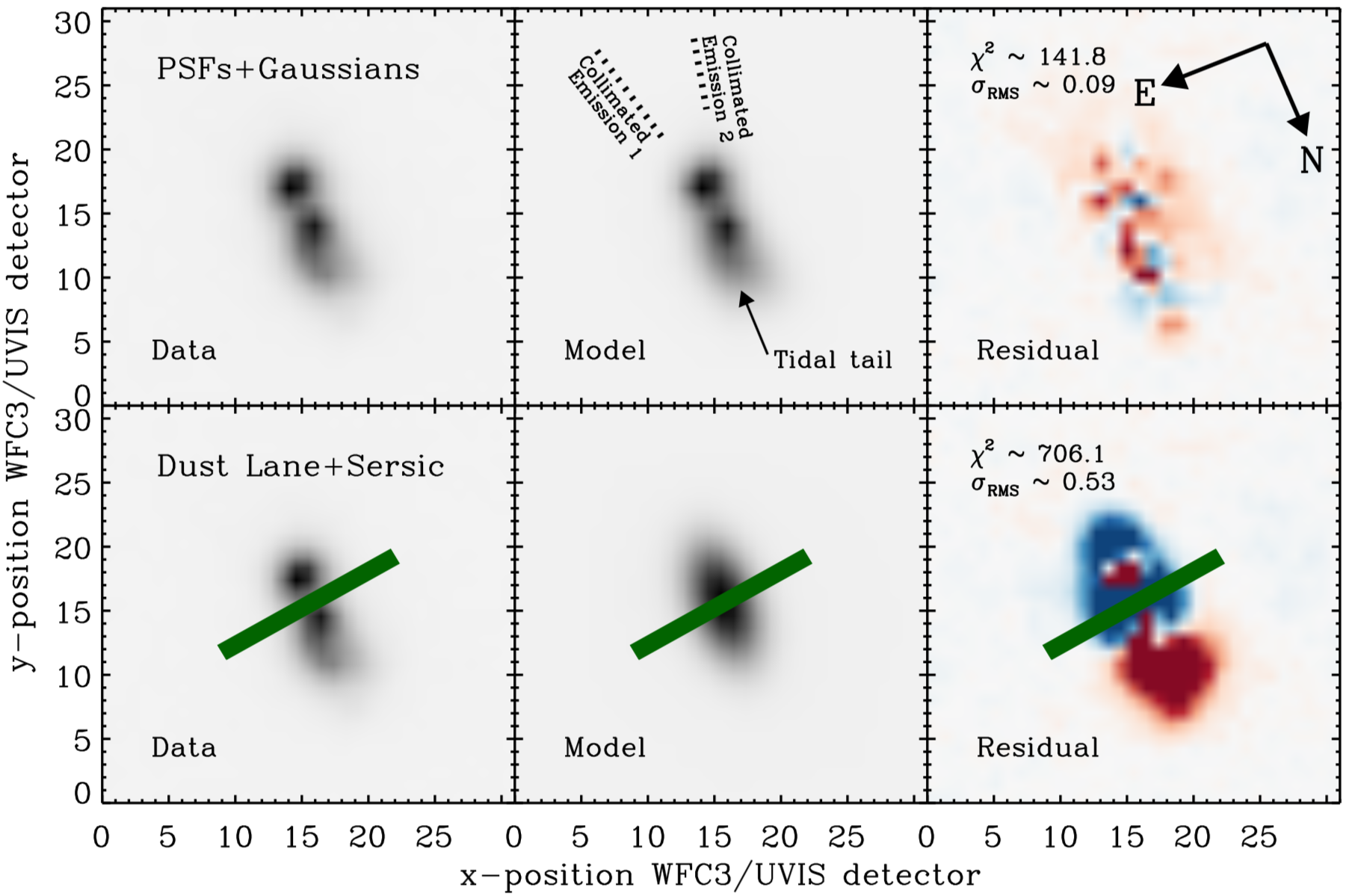}
    \caption{Morphological analysis of emission in the central 1$''$x1$''$ region of J1010+1413 using the {\sc galfit} software package. Upper: our preferred 4-parameter model (two PSFs and two extended Gaussians). The residual image (data -- model) consists of only unstructured Poisson noise with no distinct features. Lower: Sersic profile with a (presumed) dust lane (represented by a mask; green line) splitting the observed emission in two. The residuals show significant structure with a high RMS noise.}
    \label{fig:galfit}
\end{figure*}


\subsection{Chandra ACIS-S observations}
J1010+1413 was observed on January 20, 2016 with the Advanced CCD Imaging Spectrometer (ACIS) on board the NASA Chandra X-ray Observatory (Proposal:17700576; PI:Pardo). The data were reduced following standard procedures using the Chandra Interactive Analysis of Observations (CIAO) software package provided by the Chandra X-ray Center. After standard grade filtering, the total exposure in good time intervals was $\sim$24.1~ks. The native spatial resolution of ACIS-S (0.492$''$/pixel) is insufficient to resolve two distinct point sources separated by only 0.13$''$, even when using more sophisticated sub-resolution techniques, and hence, the quasars should appear as a single X-ray point source. The soft-band image (0.5--3.5~keV) is characterized by marginally extended ($r \sim 1.9''$) diffuse emission with 59$\pm$8 counts. A spectral analysis of the X-ray emission (extracted from a $2''$ aperture and grouped to 1 photon/bin using {\tt grppha}) involving a simple absorbed power law combined with a Galactic foreground absorber ($N_{\rm H} = 3.4 \times 10^{20}$cm$^{-2}$) produces a best-fit $\Gamma \sim 2.8 \pm 0.3$ and $N_{\rm H} < 10^19$cm$^{-2}$, which is inconsistent with X-ray emission arising directly from an AGN. An APEC plasma model is also found to be a reasonable description of the data with $kT\sim3.1\pm0.6$, although the soft power law model is marginally preferred ($\Delta C_{\rm stat} \sim$2.2). The best-fit X-ray flux is only $f_{\rm X,0.5-7keV} \sim 2.0\pm0.5 \times 10^{-14}$erg~s$^{-1}$~cm$^{-2}$, i.e., a low luminosity of $L_{\rm 0.5-7keV} \sim 1.5 \times 10^{42}$erg~s$^{-1}$, and is a factor $\gtrsim 650$ below the naive expectation based upon the $L_{\rm 6\mu m}$--$L_{\rm X}$ relation \citep{Chen2017}. Such a deficit in the observed X-ray emission compared with other mid-IR/optical $L_{\rm AGN}$ indicators \citep{Alexander2008,Goulding2011} suggests obscuration by heavily Compton-thick ($N_{\rm H} \sim 10^{25}$~cm$^{-2}$) gas. The additional inclusion of a second absorbed power law to assess the presence of a heavily obscured Compton-thick quasar was left unconstrained due to the small number of photons.

Given its low luminosity, spectral shape, and the fact that it is resolved, the soft X-ray emission most likely arises from star formation or quasar-produced {\it scattered} light, similar to the well-studied quasar SDSSJ1356+1026 \citep{Greene2014}. However, the observed H$\alpha$ emission ($L_{H\alpha} \sim 1.3 \times 10^{42}$erg~s$^{-1}$) allows us to place a conservative upper limit (i.e., assuming no AGN contribution to H$\alpha$) on the star-formation rate (SFR) of $\lesssim$10\Msun/yr. When coupled with the total stellar mass and the relation of \cite{Lehmer10}, we can place an upper limit on the X-ray emission from star formation of $L_X \lesssim 3 \times 10^{40}$erg~s$^{-1}$. Hence, the slightly higher observed X-ray luminosity is more suggestive of a scattered light origin.

\section{Morphological Evidence for a Pair of Accreting SMBHs}

Due to the presence of two distinct nuclear stellar cores with cospatial [OIII] emission radiating at quasar-like luminosities, our preferred interpretation is that J1010+1413 contains two rapidly accreting SMBHs close to its photometric center. Such a scenario is consistent with the very late-stage merging galaxy found from previous SDSS and Gemini-IFU data. However, even though both the F689M continuum and [OIII] images strongly suggesting two distinct, spatially-separated point sources, complicated structures and morphologies in the NLRs can make the identification of SMBH pairs ambiguous \citep{Shen2011b}. 

To better elucidate the nature of the nuclear region in J1010+1413, we performed an imaging decomposition in the F689M and [OIII] images using the {\sc galfit} package \citep{Peng10}. The {\sc galfit} analysis requires an accurate representation of the WFC3/UVIS point spread function (PSF) to convolve with the model components. We constructed the PSF using a hybrid methodology \citep{Candels11}, which combines a model of the instrument PSF (produced by the STScI software package TinyTim) with real point-like stellar objects detected in the observation. We confirmed that our PSF model produces an accurate subtraction of a point source in the WFC3 images, with no clear systematic residuals beyond the image noise. 

The [OIII] emission is not well fit using only two PSF functions; rather it requires the inclusion of two extended Gaussians, centered close to each of the point sources. Conversely, a fit using only two Gaussians produces a significantly lower quality fit ($\Delta \chi^2 = (\chi^2_{\rm 4comp}-\chi^2_{\rm 2comp}) \sim -189$) leaving residual point sources in the image, and hence, two PSF functions are also required. The best-fit solution from the 4-component model is shown in Fig.2 (upper panel), producing an excellent fit to the data; the residual is consistent with Poisson noise. Each Gaussian component is a factor $\sim$2-3 brighter than its associated PSF component; this is to be expected as NLRs are typically observed on scales of several hundred parsecs in local galaxies, and should therefore require a bright component extended beyond the PSF. Furthermore, one of the Gaussian components is marginally offset $\sim$0.1$''$ north of the northern nucleus, and is consistent with the termination of a stream of stellar material, presumably the result of the on-going merging of the two cores. This component is observed in both the F689M and [OIII] images, while a continuation of this stream is clearly visible to the west of the core in the F689M image shown in Fig.1. 

A potential alternative scenario for the observed features could be a single SMBH residing at the center of a large extended stellar bulge and NLR, split along the minor axis by a dust lane. This dust lane would fully obscure the would-be point source emission arising from a single quasar. If the gas is uniformly distributed, the [OIII] emission arising from a single SMBH would appear as an elongated component above and below the dust lane. To test this, we attempted to model both the continuum and [OIII] images using a single Sersic component with the centroid constrained to the region hidden by an artificial dust lane (represented by a {\sc galfit} mask region; Fig.2 lower panels). The resultant Sersic component has an extremely flat profile ($n\sim0.4$), similar to that of a Gaussian. However, in comparison to our best-fit model, the Sersic+dust-lane model is an extremely poor fit $\Delta \chi^2 = (\chi^2_{\rm Best}-\chi^2_{\rm DL+Sersic}) \sim -564.3$; the residual RMS noise is a factor $\gtrsim 5$ larger than our best-fit model, and the two previously determined point sources are strongly under-subtracted in the residual image. Hence, the HST data do not support a scenario involving a single SMBH hidden by a dust lane.

We further tested whether the very luminous [OIII] emission could be produced by shock heating from powerful starburst-driven winds, ignited as a result of the galaxy merger. Given either the [OIII] or 22$\mu $m luminosities, such a scenario would require SFR$\gg 10^3 $\Msun/yr in the central 500~pc region. However, this is inconsistent with the non-detection of J1010+1413 at 100$\mu$m in IRAS, and is orders of magnitude higher than the upper limit of $\lesssim$10\Msun/yr set by H$\alpha$. We thus conclude that the [OIII] and IR emission must arise from quasar activity.

\begin{figure}
    \centering
    \includegraphics[scale=0.5,width=0.45\textwidth]{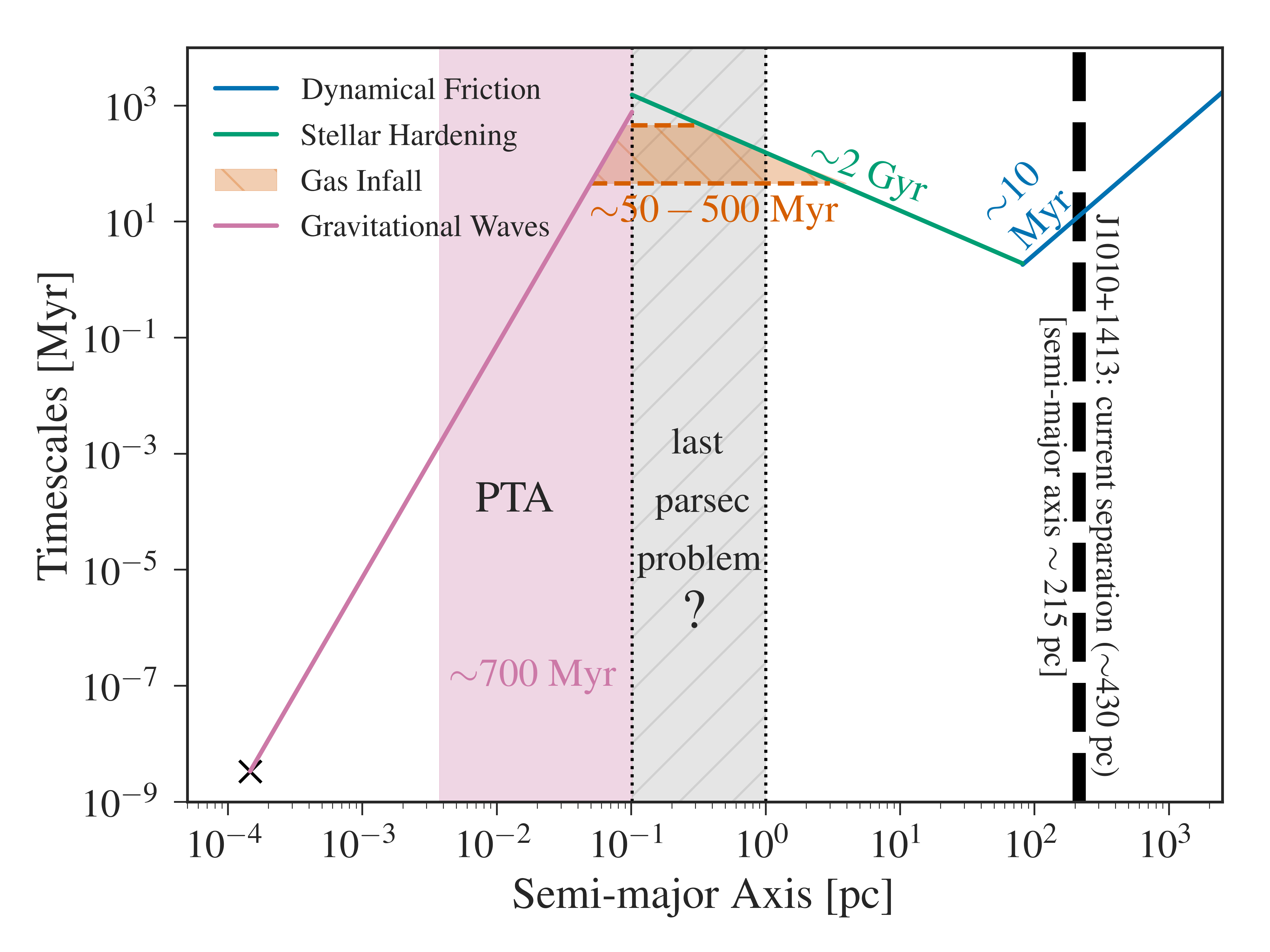}
    \caption{Dynamical timescales for J1010+1413 as a function of binary semi-major axis (assuming a circular orbit). Dynamical friction, stellar hardening, and GW emission phases are shown with blue, green, and pink lines, respectively. Current pair semi-major axis is shown with the black, dashed line. The PTA band for an object of J1010+1413’s chirp mass is indicated by the pink region.}
    \label{fig:binary_lifestory}
\end{figure}


\begin{figure}
    \centering
    \includegraphics[width=0.45\textwidth]{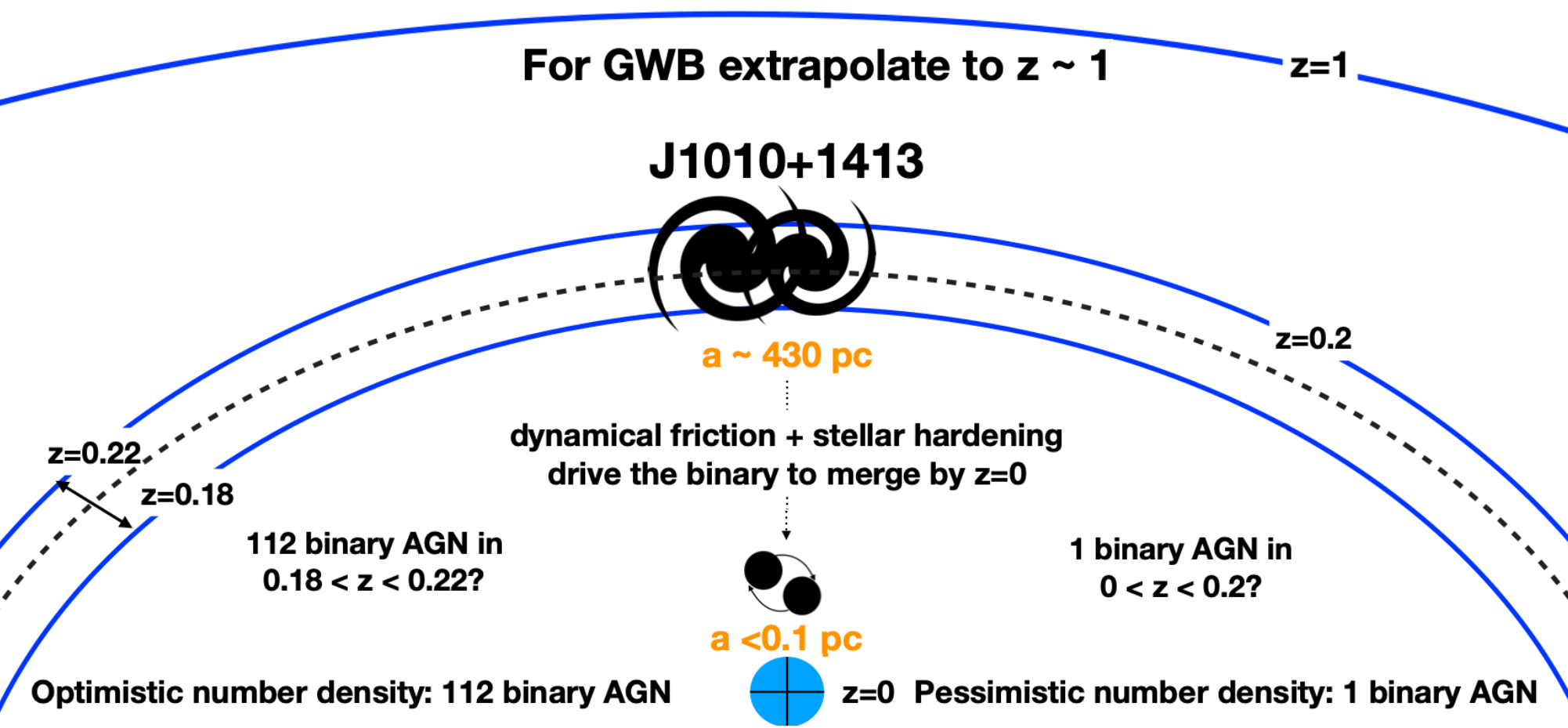}
    \caption{Schematic of GWB amplitude assumptions given the SMBH pair in J1010+1413. Dynamical friction and stellar hardening will drive SMBH pairs like J1010+1413 to merge by $z=0$ (Fig.3), implying that there is at least one local SMBH binary emitting GWs. The expectation of this one GW source is used to estimate $n_{\rm binary}(z=0)$, which is extrapolated to $z=1$ to compute a pessimistic GWB amplitude estimate, $A\sim 1.8\times 10^{-17}$. Alternately, a more optimistic $n_{\rm binary}(z=0)$ can be estimated based on the population of known luminous quasars similar to J1010+1413 within $0.18\leq z\leq 0.22$, producing $A\sim 3\times 10^{-16}$. }
    \label{fig:smbh_evo}
\end{figure}


Our {\sc galfit} simulations confirm the presence of two continuum and [OIII]-emitting cores in the nuclear region of J1010+1413. The southern nuclear region is marginally more luminous in [OIII] than the northern nucleus with $\Delta m_{\rm [OIII]} \sim 0.14$, i.e., a flux ratio of 1:1.3. Using the previously measured Balmer decrement \citep{Mullaney13} and an extinction-corrected bolometric correction \citep{Kauffmann2009}, we measure intrinsic quasar luminosities of $L_{\rm AGN,N} \sim 4.2 \times 10^{46}$ and $L_{\rm AGN,S} \sim 5.4 \times 10^{46}$~erg~s$^{-1}$ for the north and south nucleus, respectively. The combined luminosities are consistent with $L_{\rm bol}$ determined in the mid-IR. Under the presumption that both quasars are accreting at $L/L_{\rm Edd} = 1$, we are able to place a {\it minimum} mass of $4 \times 10^8$\Msun\ for each SMBH in the pair.

We conclude that when all lines of evidence are taken together -- (1) the measured mid-IR and [OIII] luminosity; (2) the morphology of the [OIII] emission; and (3) the spatial coincidence of the stellar-continuum and [OIII] point sources -- J1010+1413, in all likelihood, harbors two roughly equal-mass $M_{\rm BH} \gtrsim 4 \times 10^8$\Msun\ SMBHs separated in projection by $\sim$430pc. Further evidence for a pair of SMBHs in J1010+1413 must await on-going follow-up with JVLA, HST and ground-based adaptive optics telescopes.

\begin{figure*}
    \centering
    \includegraphics[width=\textwidth]{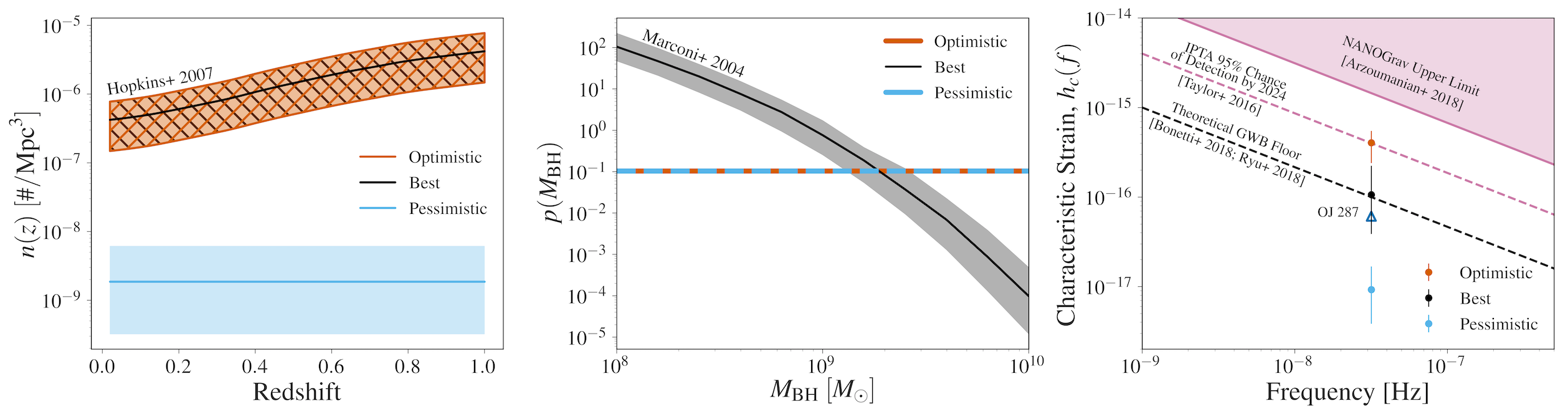}
    \caption{{\bf Left:} Number density $n(z)$ of AGN binaries pessimistically assuming J1010+1413 is the only binary AGN in SDSS to $z<0.2$ and constant number density. We assume an evolving $n(z)$ normalized to the number of J1010+1413-like systems at $z \sim 0.18$--0.22 in SDSS for our most optimistic (orange) and best-case (black) scenarios, which themselves differ by the assumed SMBH mass function (center panel). Shaded regions provide estimated uncertainties. {\bf Center:} Assumed SMBH mass functions for binary AGN. {\bf Right:} Estimated characteristic strain ($h_c$) of the GWB following our well-motivated $n(z)$ and $n(M_{\rm BH})$ assumptions. For reference, we provide $h_c$ (95\% lower limit) assuming the blazar OJ287 is a true SMBH binary \citep{Zhu2019, dlg+19}.}
    \label{fig:limits}
\end{figure*}


\section{Relevance to Gravitational Waves}

J1010+1413 is not currently emitting GWs in the PTA band. However, as we describe below, the predicted time to the PTA band for J1010+1413 is less than the lookback time to the object (Fig~3). Thus, we expect a population of objects like J1010+1413 that would be emitting GWs in the PTA band \textit{today}, Fig. \ref{fig:smbh_evo}. Using well-motivated assumptions on their number density ($n_{\rm binary}$), we can place limits on the GWB, Fig. \ref{fig:limits}.

\subsection{Coalescence predictions}

The evolution of SMBH pairs is expected to proceed through three main stages leading to coalescence: (1) dynamical friction, (2) stellar hardening, and (3) GW emission \citep{Begelman1980}. The SMBH pair in J1010+1413, currently separated by $430~\rm{pc}$, should be nearing the end of its dynamical friction phase (Fig.3). This phase has a timescale given by \citep{Binney2008}:
\begin{equation}
    t_{\rm{dyn}} = \frac{19\ \rm{Gyr}}{\ln \Lambda} \left( \frac{a}{5\ \rm{kpc}}\right)^2 \frac{\sigma}{200\ \rm{km/s}} \frac{10^8\ M_{\odot}}{M_2} \; , 
\end{equation}
where $a$ is the semi-major axis of the binary, $\sigma$ is the velocity dispersion of the stars, and $M_2$ is the mass of the lighter SMBH. We use the virial theorem with our measured values for the $M_{*}$ and $r_{\rm gal}$ to estimate the stellar velocity dispersion. We expect the SMBH pair will enter the stellar hardening phase within ${\sim} 10~\rm{Myr}$, at which time the SMBH pair's angular momentum will be transferred to slower-moving stars that pass close to the pair, decreasing the orbital separation. This stellar hardening happens on a timescale given by \citep{Sesana2015}:
\begin{equation}
    t_{\rm{sh}} = \frac{\sigma_{\rm{inf}}}{G H \rho_{\rm{inf}} a_{\star/\rm{gw}}}\; ,
\end{equation}
where $\sigma_{\rm{inf}}$ is $\sigma$ at the BH influence radius, $\rho_{\rm{inf}}$ is the mass density at this radius, $H = 15$ is a dimensionless hardening rate constant \citep[see][for more details]{Quinlan1996}, and $a_{\star/\rm{gw}}$ is the semi-major axis at which GW emission begins to dominate the decay of the orbit. We follow the usual procedure of assuming a Dehnen profile with $\gamma =1$ for the stars \citep{Dehnen1993, Sesana2015, Mingarelli2017}. However, there may be insufficient stars to eject for the pair to reach the sub-pc scale, and merge via GW emission \citep{Yu2003}. In this `final parsec problem' scenario, the time to coalescence becomes significantly longer than a Hubble time, causing the pair to stall. Without taking this effect into account, the expected hard binary timescale is $\sim 2~\rm{Gyr}$. Large quantities of gas surrounding the SMBH pair will decrease the binary's time to coalescence.For approximately equal-mass systems, the timescale for migration is about equal to the accretion timescale \citep{Gould2000}. Assuming the SMBH pair in J1010+1413 continues to accrete at $\lambda_{\rm Edd} = 0.1$--1.0, the gas accretion phase will require  \citep{Begelman1980}:
\begin{equation}
    t_{\rm{gas}} \sim \frac{M_{1}}{M_{\odot}} \left(\frac{\dot{M}}{1~M_{\odot}/\rm{yr}}\right)^{-1}~\rm{yr} \; ,
\end{equation}
where $\dot{M}$ is the accretion rate, i.e., a gas accretion timescale of $t_{\rm{gas}}\sim 100$~Myr. Thus, we expect the SMBH pair to reach sub-parsec separations within $0.1-2~\rm{Gyr}$. Once $r_{\rm sep} \lesssim 0.1~\rm{pc}$, GW emission will lead to final coalescence within ${\sim} 700~\rm{Myr}$. If merging SMBH binaries do not stall, by the present day we expect J1010+1413-like systems at $z\sim0.2$ to have coalesced, anchoring the number of possible SMBH mergers since $z \sim 0.2$.

\subsection{Contribution to the GWB}

A GWB produced by the incoherent superposition of GWs from all inspiralling SMBH binaries over cosmic history is expected to be observed by PTAs in the next few years \citep{Taylor16, kbh+17}. SMBH pairs at $z<1$ and in the $10^8 - 10^9$\Msun\ range, i.e., systems similar to J1010+1413, are expected to be the primary source population of this signal. The GWB amplitude depends strongly on the SMBH mass function, SMBH occupation fraction, and galaxy-galaxy merger rates, while the shape of the characteristic strain ($h_c$) spectrum holds clues to the final parsec problem \citep{Sampson15,Arzoumanian2016}. See \citet{Mingarelli2019} for a brief review, and \citet{btc+18} for a comprehensive one.

Here we use the existence of the SMBH pair in J1010+1413 to put limits on the space density of similar objects. 

As a lower limit, we assume that J1010+1413 is the only $>10^8$\Msun\ SMBH binary detectable in the SDSS to $z = 0.2$. This would imply that an analogous system would be merging today i.e., $n_{\rm{binary}}(z=0) = 1\times 10^{-9}\ \rm{Mpc}^{-3}$, Fig. \ref{fig:smbh_evo}. Assuming $n_{\rm{binary}}$ is \textit{constant} with both $M_{\rm BH}$ and $z$, we estimate there are $\sim 300$ binary AGN that contribute to the GWB to $z<1$.

Our most optimistic scenario assumes that J1010+1413 is representative of luminous Type-2 quasars ($L_{\rm bol} \gtrsim 10^{46}$~erg~s$^{-1}$) in merging systems that are detectable by SDSS, WISE and the NRAO VLA Sky Survey at $z \sim 0.18$--0.22. By selecting all Type-2 AGN from \cite{Mullaney13} at $z \sim 0.18$--0.22, and assuming $L/L_{\rm{Edd}} = 0.3$ (typical of SDSS quasars; \citealt{Shen2011}) and the $z=0.2$ observed fraction of AGN in merging systems ($f_{\rm{merge}} = 0.25$; \citealt{Hickox2014}), we predict there are potentially 112 binary AGN with $M_{\rm BH} > 10^8$\Msun\ in SDSS, i.e., $n_{\rm{binary}}(z=0) = 2\times 10^{-7}\ \rm{Mpc}^{-3}$. Uncertainties are estimated using a range of Eddington ratios ($L/L_{\rm{Edd}} = 0.1 - 1.0$). We predict the evolution of $n_{\rm{binary}}$ to $z=1$ by normalizing the quasar number density of \citet{Hopkins2007} to $n_{\rm{binary}}(z=0)$, providing a total of $\sim 1.2\times 10^6$ binary AGN to $z<1$.

Our more realistic (``best") scenario also accounts for the dependence of $n_{\rm binary}$ on $M_{\rm BH}$. We expect SMBHs in binaries to follow an $M_{\rm BH}$ function \citep{Marconi2004}, which corrects for the smaller number of very massive ($M_{\rm BH}>10^{10}~M_{\odot}$) SMBHs in the Universe. Our limits do not differ greatly if we instead adopt the observed quasar luminosity function and a fixed Eddington ratio. Allowing the number density to vary with $M_{\rm BH}$ and $z$ gives the same total number of binary AGN, but distributes their $M_{\rm BH}$ differently. We do not expect the SMBH mass function to vary significantly to $z<1$ in the range $M_{\rm BH} = 10^8-10^{10}$\Msun\ \citep{Merloni2008}. 

We use our three scenarios to compute the GWB $h_c$ using the \cite{Phinney2001} formalism:
\begin{equation}\label{eqn-gwb}
    h_c^2(f) = \frac{4G^{5/3}}{3\pi^{1/3}c^2} \frac{1}{f^{4/3}} \int n(z, \mathcal{M})\frac{(\mathcal{M})^{5/3}}{(1+z)^{1/3}}~dz~d\mathcal{M} \; ,
\end{equation}
where $\mathcal{M} = (1/(1+q)^2)^{3/5}M_1$ is the chirp mass of the binary, $q$ is the binary mass ratio, and $M_1$ is the primary mass. We integrate over $z=0-1$ and $\mathcal{M} = 10^{7.6}-10^{9.6}~M_{\odot}$, where this mass range corresponds to $M_{\rm{BH}} = 10^8 - 10^{10}~M_{\odot}$ and $q=1$. We note that computing the integral to $z=2$ increases our final $h_c$ estimate by a factor $\lesssim$2.

We predict a range of limits on $h_c(f={1~\rm{yr}^{-1}})$ from  $9.2\times 10^{-18}$ to $4.0\times 10^{-16}$, with a best estimate of $h_c(f={1~\rm{yr}^{-1}}) = 1.1\times 10^{-16}$. This is ${\sim} 1-10\%$ of the most recent upper limit on the GWB, $h_c(f={1~\rm{yr}^{-1}}) < 1.45\times 10^{-15}$ \citep{Arzoumanian2018}, and well within the reach of PTAs in the next decade \citep{Taylor16}.

Our predicted range for $h_c$ brackets the theoretical lower limit of ${\sim} 10^{-16}$, which would be produced by SMBH binaries stalled at the last parsec, assuming conservative $M_{\rm BH}$ estimates \citep{Sesana16, Shankar16}. These stalled SMBHs ultimately merge via many body interactions with SMBHs introduced by additional galaxy mergers \citep{Bonetti2018,Ryu2018}. There are therefore two intriguing implications for the GWB following our discovery of J1010+1413: (1) if the true GWB amplitude is marginally below the current sensitivity limits, then SMBH pairs similar to that of J1010+1413 contribute $1 - 10\%$ of the GWB signal, and thus there must be little stalling of the SMBH pairs in nature; or (2) if the GWB amplitude s lower than our predicted limits, then we would have evidence that $M_{\rm BH}$ have been overestimated and/or that nature does not have a solution to the final parsec problem. In concert with future simulations and/or the improving GWB upper-bound, the SMBH pair in J1010+1413 will anchor source population estimates, merger rates, and even the volume of the GWB. 

The identification of the SMBH pair in J1010+1413 has yielded new empirical insight into the nature of the nanohertz GWB. Our investigation has highlighted the benefit of combining quasar detections made in the mid-IR with high-resolution optical imaging to confirm the presence of a SMBH pair. However, our current estimates for $h_c$ are limited by our ability to (1) accurately measure the number of J1010+1413-like systems in the redshift slice $z\sim0.18-0.22$, and (2) constrain the evolution of such merging systems out to $z<1$ where the GWB signal may peak. The combination of sensitive large-scale surveys that are optimized for AGN detection and/or galaxy morphologies (e.g., SphereX; WFIRST) will allow the future discovery and characterization of a population of small-separation SMBH pairs, similar to J1010+1413. 

\acknowledgements The authors thank Ai-Lei Sun, Sean Johnson, Sarah Burke-Spolaor, and Xingjiang Zhu for valuable insight and discussions. We thank the anonymous referee for useful comments. ADG gratefully acknowledges support from National Aeronautics and Space Administration through Chandra Award GO6-17094X issued by the Chandra X-ray Observatory Center, which is operated by the Smithsonian Astrophysical Observatory for and on behalf of the National Aeronautics Space Administration under contract NAS8-03060, and through Hubble award HST-GO-14730.001-A from the Space Telescope Science Institute, which is operated by AURA, Inc., under NASA contract NAS 5-26555. KP acknowledges support from the National Science Foundation Graduate Research Fellowship Program  under  grant  DGE-1656466. The Flatiron Institute is supported by the Simons Foundation. 

\bibliographystyle{aasjournal}

\end{document}